\documentclass[aps,prd]{revtex4}
\usepackage{subfigure,graphics} 
\usepackage{flafter} 
\begin{document} 

\title{
Testing models of inflation with CMB non-gaussianity} 
\author{Ian G. Moss}
\email{ian.moss@ncl.ac.uk}
\author{Christopher M. Graham}
\affiliation{School of Mathematics and Statistics, University of  
Newcastle Upon Tyne, NE1 7RU, UK}

\date{\today}

%%%%%%%%%%%%%%%%%%%%%%%%%%%%%%%%%%%%%%%%%%%

\begin{abstract}
Two different predictions for the primordial curvature fluctuation bispectrum
are compared through their effects on the Cosmic Microwave Background
temperature fluctuations. The first has a local form described by a single
parameter $f_{NL}$. The second is based on a prediction from the warm
inflationary scenario, with a different dependence on wavenumber and a
parameter $f_{WI}$. New expressions are obtained for the angular bispectra of
the temperature fluctuations and for the estimators used to determine $f_{NL}$
and $f_{WI}$. The standard deviation of the estimators in an ideal experiment
is roughly $5$ times larger for $f_{WI}$ than for $f_{NL}$. Using 3 year WMAP
data gives limits $-375<f_{WI}<36.8$, but there is a possibility of detecting
a signal for $f_{WI}$ from the Planck satellite.

\end{abstract}
\pacs{PACS number(s): }

\maketitle
%%%%%%%%%%%%%%%%%%%%%%%%%%%%%%%%%%%%%%%%%%%
\section{introduction}

Inflationary models of the early universe have proved to be in spectacularly
good agreement with observations of the Cosmic Microwave Background. The
spectrum and the gaussianity of the temperature fluctuations are both
consistent with the simplest type of inflationary models. The quality of the
observational data will improve in coming years, and a future aim will be to
learn more about the nature of the inflationary era.

We shall focus here on the possible use of non-gaussianity to distinguish
between different inflationary scenarios. The dissadvantage with this approach
is that the amount of non-gaussianity produced in the simplest inflationary
models is very small
\cite{Falk:1992sf,Gangui:1993tt,Acquaviva:2002ud},
and it might not be possible to separate it from non-primordial sources of
non-gaussianity such as background sources \cite{Komatsu:2001rj} or the second
order Sachs-Wolfe effect \cite{Munshi:1995eh,Pyne:1995bs}. Attention has
therefore concentrated on inflationary models which produce a relatively large
amount of non-gaussianity and, fortunately, there are many examples.

An early example was the curvaton
mechanism, where the density fluctuations originate from the quantum
fluctuations of a light scalar field whose energy density is negligible during
inflation \cite{Lyth:2004gb,Sasaki:2006kq}. The non-gaussianity depends on the
energy density of the curvaton when it decays, and it may be larger than
secondary sources of non-gaussianity in the CMB in special cases
\cite{Komatsu:2001rj}. 
Another possibility is that
non-gaussianity can be produced if there exists a non-homogeneous reheating
mechanism after inflation, and the non-gaussianity may be comparable to
secondary sources \cite{Dvali:2003ar}. The possibility which we focus on here
is motivated by a recent result on warm inflation \cite{Moss:2007cv},
in which it is possible to have models where the primordial non-gaussianity is
significantly larger than the non-gaussianity produced by secondary sources.

The non-gaussianity can be quantified by taking moments of the density
fluctuations, especially the bispectrum $B_\zeta(k_1,k_2,k_3)$. The bispectrum
is defined by the stochastic average of a triple product of perturbation
amplitudes
\begin{equation}
\langle \zeta({\bf k}_1)\zeta({\bf k}_2)\zeta({\bf k}_3)\rangle=
(2\pi)^3B_\zeta(k_1,k_2,k_3)
\delta^3({\bf k}_1+{\bf k}_2+{\bf k}_3).\label{defB}
\end{equation}
Note that we will use the Bardeen variable $\zeta$ (the three-space curvature
on a constant density hypersurface \cite{bardeen83}) to
describe the primordial density fluctuations. The Bardeen variable has the
advantage of being constant for small amplitude density perturbations on large
length scales.

The shape of the bispectrum depends on the particular inflationary scenario. In
this paper we will concentrate on two different models for the bispectrum. The
first will be the local model,
\begin{equation}
B^{CI}_\zeta(k_1,k_2,k_3)=2f_{CI}\sum_{\rm cyc}P_\zeta(k_1)P_\zeta(k_2).
\end{equation}
where `cyc' indicates a cyclic permutation of $k_1$, $k_2$ and $k_3$ and the
primordial power spectrum $P_\zeta(k)$ is defined by
\begin{equation}
\langle \zeta({\bf k}_1)\zeta({\bf k}_2)\rangle=(2\pi)^3P_\zeta(k_1)
\delta^3({\bf k}_1+{\bf k}_2).
\end{equation}
This type of bispectrum describes the primordial fluctuations for inflationary
models with density fluctuations generated by a curvaton field
\cite{Lyth:2005fi}.
The usual parameter which is used in this case is not $f_{CI}$ but the
non-linearity parameter $f_{NL}$ of the Newtonian potential fluctuations. The
best observational limits
on the non-linearity at present comes from from the WMAP three-year data
\cite{Spergel:2006hy},
\begin{equation}
-54<f_{NL}={5\over3}f_{CI}<114\hbox{~~~~at 95\% C.L.}
\end{equation}
The Planck satellite observations have a predicted sensitivity limit of around
$|f_{NL}|\sim5$  \cite{Komatsu:2001rj}.

The second model is motivated by the non-gaussianity
arising from the warm inflationary scenario \cite{Moss:2007cv}. In warm
inflation there is a significant amount of radiation present during
inflation and the density fluctuations arise from thermal, rather than quantum
fluctuations \cite{Moss85,berera95}. The bispectrum takes the form
\begin{equation}
B^{WI}_\zeta(k_1,k_2,k_3)=f_{WI}\sum_{\rm cyc}(k_1^{-2}+k_2^{-2})
P_\zeta(k_1)P_\zeta(k_2){\bf k_1}\cdot{\bf k_2}.
\end{equation}
A similar bispectrum appeared in \cite{Liguori:2005rj} as part of a
longer expression describing second order perturbative effects following the
usual inflationary scenario, but there $f_{WI}\approx 3$, and here we consider
arbitrary values of $f_{WI}$. We shall consider the prospects for testing this
model using CMB data from WMAP and PLANCK experiments. For other forms of
bispectrum, see refs. 
\cite{Creminelli:2003iq,Creminelli:2005hu,Smith:2006ud}. 

\section{The angular bispectrum}

We follow ref. \cite{Liguori:2005rj} and express the primordial bispectrum as a
Legendre polynomial expansion,
\begin{equation}
B_\zeta(k_1,k_2,k_3)=\sum_{\rm cyc}\sum_lf_l(k_1,k_2)
P_\zeta(k_1)P_\zeta(k_2)P_l(\hat{\bf k_1}\cdot\hat{\bf k_2}).\label{bgen}
\end{equation}
Suppose that the $f_l(k_1,k_2)$ are homogeneous functions of momenta, then
without any loss of generality, we can write
\begin{equation}
B_\zeta(k_1,k_2,k_3)=\sum_{n,l}f_{nl}B^{nl}(k_1,k_2,k_3),\label{bf}
\end{equation}
where
\begin{equation}
B^{nl}(k_1,k_2,k_3)=\sum_{\rm cyc}(k_1^nk_2^{-n}+k_2^nk_1^{-n})
P_\zeta(k_1)P_\zeta(k_2)P_l(\hat{\bf k_1}\cdot\hat{\bf k_2}).
\end{equation}

The CMB temperature fluctuations have an angle-averaged bispectrum
$B_{l_1l_2l_3}$, defined by
\begin{equation}
B_{l_1l_2l_3}=\sum_{m_1m_2m_3}\pmatrix{l_1&l_2&l_3\cr m_1&m_2&m_3}
\langle a_{l_1m_1}a_{l_2m_2}a_{l_3m_3}\rangle\label{defang}
\end{equation}
where the matrix is the Wigner $3j$ symbol and temperature fluctuations have
been expanded in spherical harmonics,
\begin{equation}
{\Delta T(\hat{\bf n})\over T}=
\sum_{lm}a_{lm}Y_{lm}(\hat{\bf n}).
\end{equation} 
The coefficients in the spherical harmonic expansion are related to the
curvature fluctuations by
\begin{equation}
a_{lm}=4\pi(-i)^l\int {d^3k\over (2\pi)^3}
\zeta({\bf k})g_l({\bf k})Y_{lm}^*(\hat{\bf k})\label{defalm}
\end{equation}
(Note that many references on the microwave background use the transfer
function from the Newtonian potential fluctuation. We use the curvature
fluctuation, which is currently used by CMBFAST software
\cite{Seljak:1996is}.) The angle-averaged bispectrum for the curvature
bispectrum (\ref{bf}) will decompose into a series of terms,
\begin{equation}
B_{l_1l_2l_3}=\sum_{n,l}f_{nl}B^{nl}_{l_1l_2l_3}
\end{equation}
where, from eqs. (\ref{defB}),  (\ref{defang}) and (\ref{defalm}),
\begin{eqnarray}
B^{nl}_{l_1l_2l_3}&=&
(-i)^{l_1+l_2+l_3}{1\over \pi^3}\sum_{m_1m_2m_3}
\pmatrix{l_1&l_2&l_3\cr m_1&m_2&m_3}\times\nonumber\\
&&\int \prod_{i=1}^3\left(d^3k_i\,g_{l_i}(k_i)
Y^*_{l_im_i}(\hat{\bf k}_i)\right)
B^{nl}(k_1,k_2,k_3)\,
\delta({\bf k}_1+{\bf k}_2+{\bf k}_3)
\end{eqnarray}
This can be simplified by expanding the delta and Legendre functions in terms
of spherical harmonics and eliminating the momenta. After using Wigner $3-j$
symbol identities (see ref \cite{brink}), the result is that 
\cite{Liguori:2005rj}
\begin{equation}
B^{nl}_{l_1l_2l_3}=
\sum_{\rm perm}\sum_{l_1'l_2'}
c(l_1'l_2'l;l_1l_2l_3)
\int_0^\infty b^n_{l_1{l_1}'}(r)b^{-n}_{l_2{l_2}'}(r)b^{NL}_{l_3}(r)r^2dr,
\label{bc}
\end{equation}
where `perm' denotes a permutation of $l_1$, $l_2$ and $l_3$. The dependence on
the transfer function is contained in two new functions
\begin{eqnarray}
b^n_{ll'}&=&{2\over\pi}\int_0^\infty
dk\,k^{n+2}P_\zeta(k)g_l(k)j_{l'}(kr)\\
b^{NL}_l&=&{2\over\pi}\int_0^\infty
dk\,k^2g_l(k)j_{l}(kr),\label{defnl}
\end{eqnarray}
where our notation is based on ref \cite{Komatsu:2001rj}. (Most recent papers
have used Greek letters for the $b_l$'s, e.g. $\alpha_l$ for $b^{NL}_l$, but
this notation does not extend very well due a shortage of usable Greek
letters.)

The coefficients $c(l_1'l_2'l;l_1l_2l_3)$ were first obtained by Komatsu
\cite{Komatsu:2002db}. They are related to Wigner $6j$ symbols,
\begin{eqnarray}
c(l_1'l_2'l;l_1l_2l_3)&=&
\sqrt{(2l_1+1)(2l_2+1)(2l_3+1)\over 4\pi}(2l_1'+1)(2l_2'+1)
i^{2l+2l_3+l_1'+l_2'-l_1-l_2}\times\nonumber\\
&&\pmatrix{l_1'&l_2'&l_3\cr0&0&0}
\pmatrix{l_1'&l_1&l\cr0&0&0}
\pmatrix{l_2'&l_2&l\cr0&0&0}
\left\{\matrix{l_2&l_1&l_3\cr l_1'&l_2'&l}\right\}.
\end{eqnarray}
The $6j$ symbols vanish unless the $l_i'$ lie in the range $l_i\pm l$, and
therefore this expansion is most useful for small values of $l$. In such
situations, eq. (\ref{bc}) combined with a good Wigner $6j$ evaluation code
can become an efficient way to evaluate the angle-averaged bispectrum.

There is a useful normalisation condition on the coefficients,
\begin{equation}
\sum_{l_1'l_2'}(-i)^{l_1'+l_2'-l_1-l_2}
c(l_1'l_2'l;l_1l_2l_3)=
\sqrt{(2l_1+1)(2l_2+1)(2l_3+1)\over 4\pi}
\pmatrix{l_1&l_2&l_3\cr0&0&0}.
\end{equation}

\subsection{Local models}

The local bispectrum corresponds to the model with $l=n=0$, which we will
denote by the label `CI'. In this case $l_i'=l_i$, $i=1,2$, and the
coefficients $c(l_1'l_2'l;l_1l_2l_3)$ reduce to
\begin{equation}
c(l_1l_20;l_1l_2l_3)=\sqrt{(2l_1+1)(2l_2+1)(2l_3+1)\over 4\pi}
\pmatrix{l_1&l_2&l_3\cr0&0&0}.\label{defc}
\end{equation}
The angle-averaged bispectrum is simply
\begin{equation}
B^{CI}_{l_1l_2l_3}=
\sum_{\rm perm}
c(l_1l_20;l_1l_2l_3)
\int_0^\infty b^{CI}_{l_1}(r)b^{CI}_{l_2}(r)b^{NL}_{l_3}(r)r^2dr,\label{Bci}
\end{equation}
where we define
\begin{equation}
b_l^{CI}(r)={2\over\pi}\int_0^\infty
dk\,k^2P_\zeta(k)g_l(k)j_{l}(kr).\label{defci}
\end{equation}

\begin{center}
\begin{figure}[ht]
\scalebox{1.0}{\includegraphics{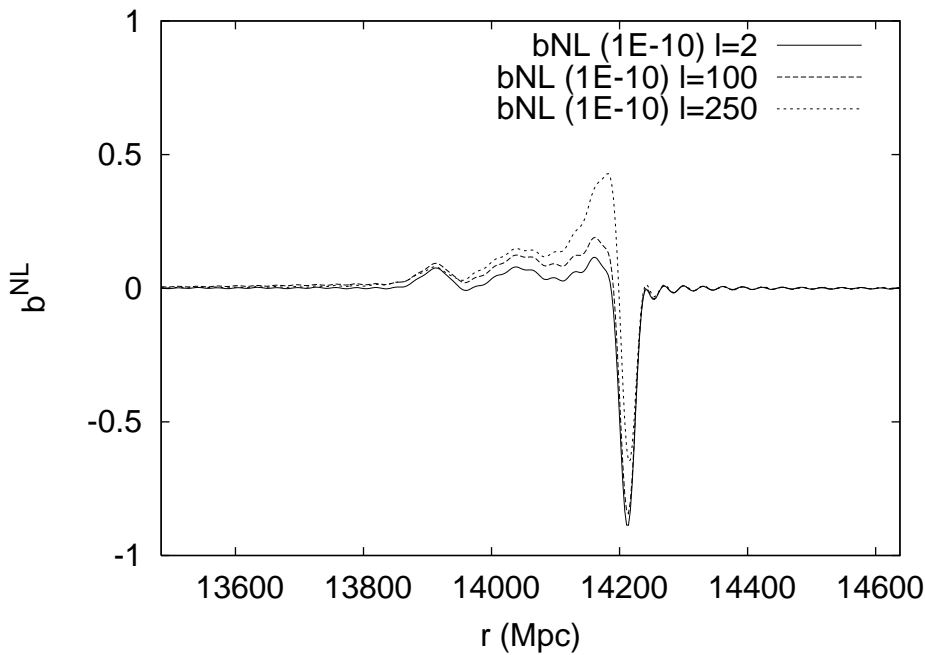}}
\caption{The function $b_l^{NL}(r)$ is plotted as a function of $r$ in Mpc for
$l=2$, $l=100$ and $l=250$. The cosmological model is the WMAP $\Lambda$CDM
plus $n=1$ model, which has $\Omega_\Lambda=0.788$, $\Omega_m=0.212$ and
$h=0.778$. With these parameters, $r_*=1.41$ Gpc and $r_d=323$ Mpc.}
\label{plotNL}
\end{figure}
\end{center}

The functions $b_l^{NL}(r)$ and $b_l^{CI}(r)$ are evaluated using CMBFAST or
related software \cite{Seljak:1996is}. Some care has to be exercised in the
evaluation of $b_l^{NL}(r)$. It is important that the time evolution of the
density perturbations in CMBFAST starts early in order to catch the large $k$
behaviour of the transfer function correctly. The function, $b_l^{NL}(r)$ has
a large spike not far from $r=r_*$, where
\begin{equation}
r_*=c\int_{t_d}^{t_0}{a(t_0)\over a(t)}dt
\end{equation}
is the speed of light times the conformal time difference between decoupling
$t_d$ and the present $t_0$. This is shown in figure \ref{plotNL}. The
resolution in $r$ has
to be very high to capture this accurately, but this incurs a large time
penalty when evaluating the angle averaged bispectrum for many different
values of $l$. We get around this problem by evaluating $b_l^{NL}(r)$ with a
high resolution (3200 values of $r$ in the range $r_*\pm r_d$) and
then averaging to produce a function $\bar b_l^{NL}(r)$ with a lower
resolution. The integral over $r$  can be obtained accurately using 
$\bar b_l^{NL}(r)$ because the other terms in the integrand vary comparatively
slowly with $r$. More sophisticated methods of reducing the number of points
in the $r$ integral are described in ref \cite{Smith:2006ud}

\subsection{Non-local models}

The second model which we consider in detail is the model with $l=n=1$, which
we will denote by the label `WI'. The coefficients $c(l_1'l_2'1;l_1l_2l_3)$
can be found using explicit formulae for the $6j$ symbols, given in ref
\cite{brink}, for example. They can be expressed conveniently in terms of
Edmond's angle $\theta$, defined by
\begin{equation}
\cos\theta=2{l_1(l_1+1)+l_2(l_2+1)-l_3(l_3+1)\over(2l_1+1)(2l_2+1)}
\label{defe}
\end{equation}
This angle is an interior angle of the simplex with sides corresponding to the
entries in the $6j$ symbol. The non-vanishing coefficients are given by
\begin{eqnarray}
{c(l_1\pm1,l_2\pm1,1;l_1,l_2,l_3)\over c(l_1l_20;l_1l_2l_3)}&=&
-{1\over 4}\left(\cos\theta+1\pm{1\over 2l_1+1}
\pm{1\over 2l_2+1}+{1\over (2l_1+1)(2l_2+1)}\right)\\
{c(l_1\pm1,l_2\mp1,1;l_1,l_2,l_3)\over c(l_1l_20;l_1l_2l_3)}&=&
-{1\over 4}\left(\cos\theta-1\mp{1\over 2l_1+1}
\pm{1\over 2l_2+1}+{1\over (2l_1+1)(2l_2+1)}\right).
\end{eqnarray}

\begin{center}
\begin{figure}[htb]
\scalebox{1.0}{\includegraphics{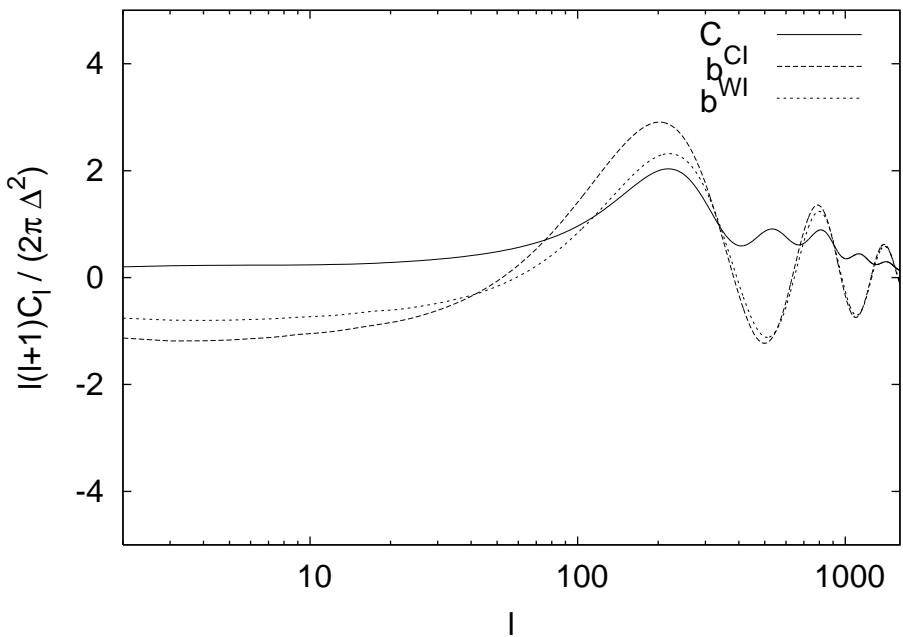}}
\caption{The $C_l$, $b_l^{CI}(r_*)$ and $b_l^{WI}(r_*)$ are plotted as a
function of $l$. The $C_l$ and $b_l^{CI}(r_*)$ have been scaled by
$l(l+1)/(2\pi\Delta^2)$, where $\Delta^2=2.5\times10^{-9}$, and the 
$b_l^{WI}(r_*)$ by an additional 
$(l-1)(l+2)/r_*^2$. The cosmological model is the WMAP $\Lambda$CDM plus $n=1$
model, which has $\Omega_\Lambda=0.788$, $\Omega_m=0.212$ and $h=0.778$. With
these parameters, $r_*=1.41$ Gpc and $r_d=323$ Mpc.}
\label{plotCl}
\end{figure}
\end{center}

The terms depending on $r$ can be simplified by using Bessel function
identities,
\begin{eqnarray}
b^1_{ll+1}&=&{l\over r}b_l^{CI}(r)-b_l^{CI\prime}(r)\\
b^1_{ll-1}&=&{l+1\over r}b_l^{CI}(r)+b_l^{CI\prime}(r)\\
b^{-1}_{ll+1}&=&{l\over r}b_l^{WI}(r)-b_l^{WI\prime}(r)\\
b^{-1}_{ll-1}&=&{l+1\over r}b_l^{WI}(r)+b_l^{WI\prime}(r),
\end{eqnarray}
where the prime denotes a derivative with respect to $r$ and $b_l^{CI}(r)$ is
given in eq. (\ref{defci}). We have introduced a new function $b_l^{WI}(r)$,
defined by
\begin{equation}
b_l^{WI}(r)={2\over\pi}\int_0^\infty
dk\,P_\zeta(k)g_l(k)j_{l}(kr).
\end{equation}
This is plotted in figure \ref{plotCl}, together with the angular power
spectrum $C_l$, defined by,
\begin{equation}
C_l={2\over\pi}\int\,dk\,k^2\,P_\zeta(k)\,g_l(k)^2.\label{defcl}
\end{equation}

The angle-averaged bispectrum $B^{WI}_{l_1l_1l_3}$ is given by substituting the
above functions into eq. (\ref{bc}),
\begin{eqnarray}
B^{WI}_{l_1l_1l_3}&=&
-\sum_{\rm perm}c(l_1l_20;l_1l_2l_3)\times\nonumber\\
&&\int_0^\infty\,dr
\left( \left(l_1+{1\over 2}\right)\left(l_2+{1\over 2}\right)
b_{l_1}^{CI}(r)b_{l_2}^{WI}(r)\cos\theta+
b_{l_1}^{CI\prime}(r)b_{l_2}^{WI\prime}(r)r^2
\right)b^{NL}_{l_3}(r),\label{Bwi}
\end{eqnarray}
where $c(l_1l_20;l_1l_2l_3)$ is defined in eq. (\ref{defc}), $b^{NL}_{l}(r)$ is
defined in eq (\ref{defnl}) and $b_{l}^{CI}(r)$ in eq. (\ref{defci}). The
radial functions and their derivatives can be evaluated using CMBFAST.

\section{Model testing}

\begin{center}
\begin{figure}[htb]
\scalebox{1.0}{\includegraphics{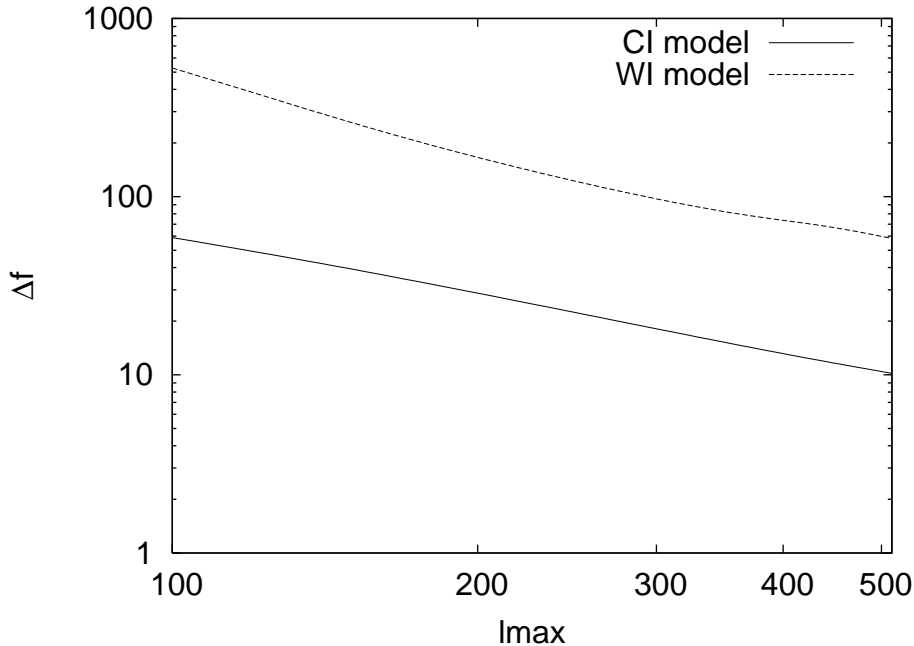}}
\caption{The theoretical standard deviation $\Delta f$ for an ideal experiment
is plotted as a function of the maximum value of $l$ for the two models of
non-gaussianity discussed in the text. The cosmological model is the WMAP
$\Lambda$CDM plus $n=1$ model, which has $\Omega_\Lambda=0.788$,
$\Omega_m=0.212$ and $h=0.778$.}
\label{plotDf}
\end{figure}
\end{center}

We shall use Maximum likelyhood methods methods to consider the sensitivity of
`ideal' experiments for detecting the non-local model of the bispectrum which
we introduced in the previous section, and we shall consider how well this
model can be distinguished from the local model. For our purposes, an ideal
experiment is one in which there is no noise in the detector and no secondary
sources of non-gaussianity.

For a single model, the log-likelyhood function 
$-\chi^2(B^{\rm obs}_{l_1l_2l_3},f)$ is a function of the nongaussianity
parameter $f$ for a set of observations $B^{\rm obs}_{l_1l_2l_3}$. For an
ideal experiment, we take the simple anzatz 
\begin{equation}
\chi^2={1\over 6}\sum_{l_1l_2l_3}{
(B^{\rm obs}_{l_1l_2l_3}-f B_{l_1l_2l_3})^2
\over C_{l_1}C_{l_2}C_{l_3}},
\end{equation}
where $C_l$ is the angular power spectrum and $C_{l_1}C_{l_2}C_{l_3}$ is the
theoretical variance of the bispectrum when the non-gaussianity is small.
The sumations extend to a maximum $l$ value $l_{max}$ which represents the
limiting resolution of the experiment.

This form of likelyhood function is associated with an estimator
\begin{equation}
\hat f=
{1\over 6F}\sum_{l_1l_2l_3} {
B^{\rm obs}_{l_1l_2l_3}B_{l_1l_2l_3}\over C_{l_1}C_{l_2}C_{l_3}}
\end{equation}
where the normalisation factor $F$ is the Fisher information,
\begin{equation}
F={1\over 6}\sum_{l_1l_2l_3} 
{(B_{l_1l_2l_3})^2\over C_{l_1}C_{l_2}C_{l_3}}.\label{deff}
\end{equation}
For an ideal experiment, the mean of $\hat f$ is the value of the parameter $f$
and the variance of $\hat f$ is $F^{-1}$. 

The standard deviation $\Delta f=F^{-1/2}$ has been plotted in figure
\ref{plotDf}, using the values of $B_{l_1l_2l_3}$ discussed in the previous
section. The WMAP $\Lambda$CDM plus $n=1$ cosmological model has been used to
define the cosmological parameters. The standard deviation for the local model
is in good agreement with previous calculations, even when they use different
cosmological parameters \cite{Creminelli:2005hu}.

\begin{center}
\begin{figure}[htb]
\scalebox{1.0}{\includegraphics{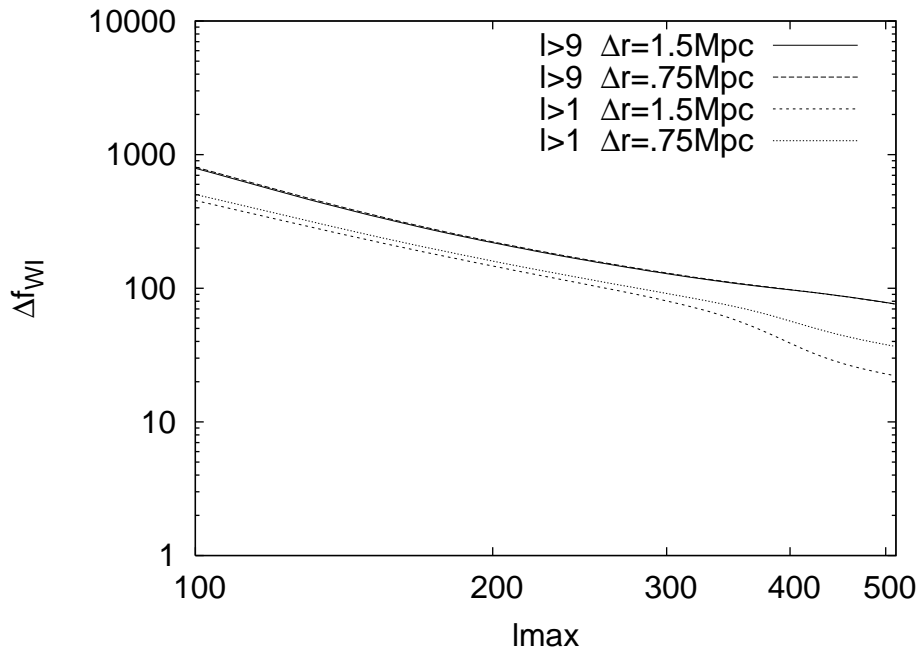}}
\caption{The theoretical standard deviation $\Delta f$ for the WI model
is plotted as a function of the maximum value of $l$ with different resolutions
in the radial integrals and different lower limits on the values of $l$. The
cosmological model is the WMAP $\Lambda$CDM plus $n=1$ model, which has
$\Omega_\Lambda=0.788$, $\Omega_m=0.212$ and $h=0.778$.}
\label{plotRv}
\end{figure}
\end{center}

The calculation of the standard deviation for the WI model with $l_{max}=512$
takes about 30 minutes on a 2GHz Intel Core Duo processor. The time is taken
up by the high resolution used to evaluate the radial integrals when we get to
values of $l\sim 400$. The high resolution is needed mostly for the `squeezed
triangles', where one value of $l$
in the summand is far smaller than the other two. This fact shows up clearly
in figure \ref{plotRv}, where the
plots show that the dependence on resolution dissappears when low multipoles
have been omitted from the sum. Our resolution
is adequate for the maximum value $l=300$ in the next section. For the Planck
experiment we would need to push the limit up to $l=2000$ or so. In this
higher range, an increase in the standard deviation arising from dropping low
multipoles may well be acceptable as a trade-off for having to perform the
calculations with many values of $r$. For example, the resolution used here is
already sufficient to obtain the variance of the estimator for the WI model
from modes in the range $9<l<1800$, and it comes out to be $\Delta
f_{WI}=15.9$. This is an upper bound on the variance of the estimator which
uses all of the modes $1<l<1800$. 

\begin{center}
\begin{figure}[ht]
\scalebox{1.0}{\includegraphics{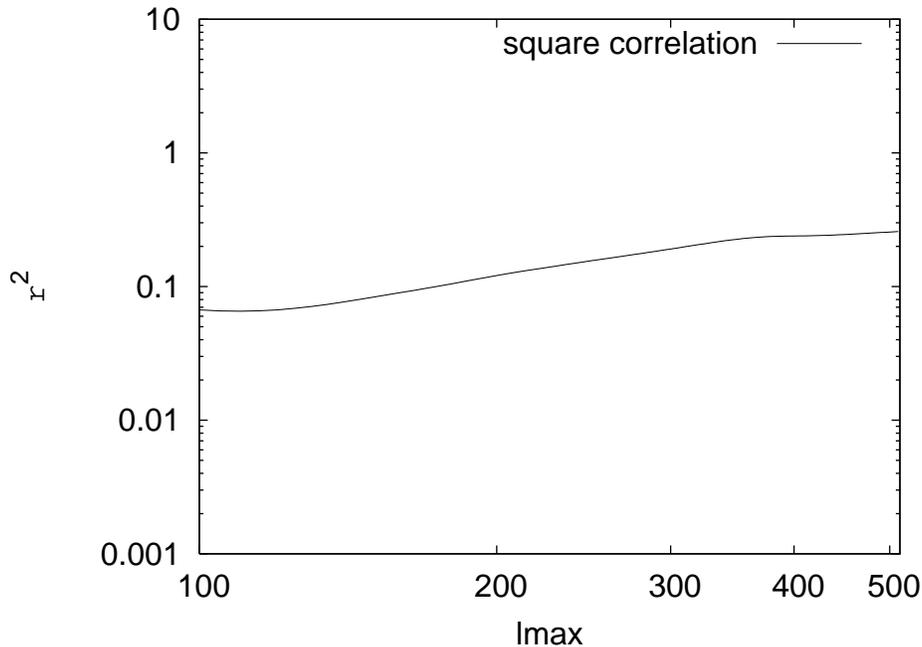}}
\caption{The square of the correlation $r$ for an ideal experiment is plotted
as a function of the maximum value of $l$ for the two models of
non-gaussianity discussed in the text. The cosmological model is the WMAP
$\Lambda$CDM plus $n=1$ model, which has $\Omega_\Lambda=0.788$,
$\Omega_m=0.212$ and $h=0.778$.}
\label{plotco}
\end{figure}
\end{center}

The observations can also be used to distinguish between two different models 
$B^i_{l_1l_2l_3}$, $i=1,2,\dots$. The Fisher information is now a matrix,
\begin{equation}
F_{ij}={1\over 6}\sum_{l_1l_2l_3} {
B^i_{l_1l_2l_3}B^j_{l_1l_2l_3}\over C_{l_1}C_{l_2}C_{l_3}}.
\end{equation}
The separability of two different models is associated with a low value for the
correlation
\begin{equation}
r=-{F_{12}\over\sqrt{F_{11}F_{22}}}.
\end{equation}
(This quantity was called the `cosine between two distributions' in ref.
\cite{Babich:2004gb}). If two bispectra have good separability, then the
failure to detect any non-gaussianity for one bispectrum does not rule out the
possibility of detection of non-gaussianity in the other bispectrum. The square
of the correlation for the $CI$ and $WI$ models has been plotted in
figure \ref{plotco}. At $l=300$, the square of the correlation is about 10\%,
which is quite low but not low enough to separate the two models definitively
using this statistic.

\section{Non-gaussianity tests on CMB data}

The estimator for the non-gaussianity of model $i$ in an ideal experiment was
given in the previous section,
\begin{equation}
\hat f^{\rm ideal}_i=
{1\over 6F_{ii}}\sum_{l_1l_2l_3} 
{B^{\rm obs}_{l_1l_2l_3}B^i_{l_1l_2l_3}\over C_{l_1}C_{l_2}C_{l_3}},
\end{equation}
where the observed angle-averaged bispectrum is related to the spherical
harmonic coefficients by
\begin{equation}
B^{\rm obs}_{l_1l_2l_3}=
\sum_{m_1m_2m_2}\pmatrix{l_1&l_2&l_3\cr m_1&m_2&m_3}
a_{l_1m_1}a_{l_2m_2}a_{l_3m_3}.
\end{equation}
We can make use use the fact that the theoretical bispectrum eq. (\ref{Bci}) or
eq. (\ref{Bwi}) consist of a small number of terms which factorise to devise
an efficient method for evaluating the estimators based on ref.
\cite{Komatsu:2003iq}.

The Wigner $3j$ symbol is first replaced by the integral
\begin{equation}
\pmatrix{l_1&l_2&l_3\cr m_1&m_2&m_3}=
c(l_1l_20;l_1l_2l_3)^{-1}
\int d^2\hat {\bf n}\,
Y_{l_1m_1}(\hat {\bf n})Y_{l_2m_2}(\hat {\bf n})
Y_{l_3m_3}(\hat {\bf n}),
\end{equation}
where $c(l_1l_20;l_1l_2l_3)$ is defined in eq. (\ref{defc}), to give
\begin{equation}
\hat f^{\rm ideal}_i=
{1\over 6F_{ii}}\sum_{l_im_i} c(l_1l_20;l_1l_2l_3)^{-1}
\int d^2\hat {\bf n}\prod_{i=1}^3
\left(a_{l_im_i}Y_{l_im_i}(\hat {\bf n})\right)
{B^i_{l_1l_2l_3}\over C_{l_1}C_{l_2}C_{l_3}}.\label{fsum}
\end{equation}
Now we define a radial sequence of filtered sky-maps $\sigma(r,\hat {\bf n})$
by
\begin{eqnarray}
\sigma_{CI}(r,\hat{\bf n})&=&
\sum_{l,m}{b_l^{CI}(r)\over C_l}a_{lm}Y_{lm}(\hat{\bf n})\\
\sigma_{WI}(r,\hat{\bf n})&=&
\sum_{l,m}{b_l^{WI}(r)\over C_l}a_{lm}Y_{lm}(\hat{\bf n})\\
\sigma_{NL}(r,\hat{\bf n})&=&
\sum_{l,m}{b_l^{NL}(r)\over C_l}a_{lm}Y_{lm}(\hat{\bf n}).
\end{eqnarray}

For the local model, $B^{CI}_{l_1l_2l_3}$ is given by eq. (\ref{Bci}). After
interchanging the sum and the integrals in eq. (\ref{fsum}),
\begin{equation}
\hat f^{\rm ideal}_{CI}=
{1\over F_{CI}}\int d^2\hat {\bf n}\,r^2dr\,
\sigma_{CI}\sigma_{CI}\sigma_{NL}\label{eCI}
\end{equation}
This can be evaluated easily with the HEALPix \footnote{See the HEALPix website
http://www.eso.org/science/healpix} software package for analysing CMB data.

The second model has angle averaged bispectrum given by eq. (\ref{Bwi}). The
$l(l+1)$ terms in Edmond's angle $\theta$ (\ref{defe}) can be replaced by the
angular Laplacian $D^2$, defined by
\begin{equation}
D^2 Y_{lm}(\hat{\bf n})=-l(l+1)Y_{lm}(\hat{\bf n}).
\end{equation}
The estimator becomes
\begin{equation}
\hat f^{\rm ideal}_{WI}=
{1\over 2F_{WI}}\int d^2\hat {\bf n}\,dr\,
\left\{
(D^2\sigma_{CI})\sigma_{WI}\sigma_{NL}
+\sigma_{CI}(D^2\sigma_{WI})\sigma_{NL}
-\sigma_{CI}\sigma_{WI}(D^2\sigma_{NL})
-2\sigma'_{CI}\sigma'_{WI}\sigma_{NL}r^2
\right\}.
\end{equation}
We can regard this as a volume integral and use the divergence theorem,
\begin{equation}
\hat f^{\rm ideal}_{WI}=
-{1\over F_{WI}}\int d^3x\,
(\nabla\sigma_{CI})\cdot(\nabla\sigma_{WI})\sigma_{NL},\label{eWI}
\end{equation}
where $\nabla$ is the three dimensional gradient operator. This expression can
be evaluated at high speed using the subroutines for handling angular
derivatives of CMB data in the HEALPix software package. 

\subsection{Real experiments}

Galactic and extragalactic sources contaminate the cosmological microwave
signal and have to be masked out, leaving us with a set of spherical
coefficients $a_{lm}$ of the masked sky. The statistical properties of these
coefficients are affected by the masking procedure. An important change
appears in the theoretical covariance matrix
\begin{equation}
C_{l_1m_1l_2m_2}=\langle a_{l_1m_1}a_{l_2m_2}\rangle,
\end{equation}
which is no longer diagonal. Detector noise is also an important consideration.
The detector noise is not uniform over the sky and the noise in the microwave
intensity has a non-trivial covariance matrix.

Creminelli et al. \cite{Creminelli:2005hu} have introduced an improved
estimator which should be close to optimal for the CMB observations. Their
estimator
\begin{equation}
\hat f_i=\hat f_i^{\rm ideal}+\hat f_i^{\rm lin},
\end{equation}
where the linear piece
\begin{equation}
\hat f_i^{\rm lin}=-{1\over 2N}\sum_{l_1l_2l_3} 
{B^i_{l_1l_2l_3}\over C_{l_1}C_{l_2}C_{l_3}}C_{l_1m_1l_2m_2}a_{l_3m_3}.
\end{equation}
The normalisation factor $N=F_{ii}f_{\rm sky}$, where $f_{\rm sky}$ is the
fraction of the sky which remains after the mask is applied.

The linear part of the estimator can be expressed in terms of filtered maps by
following the same steps as in the beginning of this section. We shall
concentrate here on the effects of the mask. To a reasonable approximation, it
can be shown that
\begin{equation}
\sum_{l_1l_2m_1m_2}b_{l_1}b_{l_2}Y_{l_1m_1}(\hat {\bf n})C_{l_1m_1l_2m_2}
Y_{l_1m_1}(\hat {\bf n})\approx
M(\hat{\bf n})\sum_l{2l+1\over 4\pi}b_l^2C_l
\end{equation}
where $b_l$ is any slowly varying function of $l$ and the mask function 
$M(\hat{\bf n})=0$ in the region being masked out and $1$ otherwise. The
details are given in the appendix. Using this approximation, the linear parts
of the estimators become
\begin{eqnarray}
\hat f^{\rm lin}_{CI}&=&
-{1\over N_{CI}}\int d^2\hat {\bf n}\,r^2dr\,M
\left( 2S_{CINL}\sigma_{CI}+S_{CICI}\sigma_{NL}\right)\\
\hat f^{\rm lin}_{WI}&=&
{1\over N_{WI}}\int d^2\hat {\bf n}\,r^2dr\,M
\left( \tilde S_{CIWI}\sigma_{NL}+
S^{(2)}_{CIWI}\sigma_{NL}+S^{(1)}_{CINL}\sigma'_{WI}+
S^{(1)}_{WINL}\sigma'_{CI}\right)
\end{eqnarray}
where the radial functions $S_{AB}(r)$, $A$, $B=CI$, $WI$ or $NL$, are defined
by
\begin{eqnarray}
S_{AB}(r)&=&\sum_l\,{2l+1\over 4\pi}{b_l^A(r)b_l^B(r)\over C_l}\\
\tilde S_{AB}(r)&=&\sum_l\,l(l+1){2l+1\over 4\pi}{b_l^A(r)b_l^B(r)\over C_l}\\
S^{(1)}_{AB}(r)&=&\sum_l\,{2l+1\over 4\pi}{b_l^A(r)'b_l^B(r)\over C_l}\\
S^{(2)}_{AB}(r)&=&\sum_l\,{2l+1\over 4\pi}{b_l^A(r)'b_l^B(r)'\over C_l}.
\end{eqnarray}
The linear parts should be combined with the other terms (\ref{eCI}) and
(\ref{eWI}).

Crenelli et al. \cite{Creminelli:2005hu} have argued that since the
contributions from the linear terms depend on angular averages of the filtered
maps, they can be reduced in size by arranging that the average value of the
data outside the mask is zero. However, we have found that the contributions
from the linear terms are important and they need to be included. 

\subsection{WMAP data}

We shall consider the non-gaussianity estimators for the 3-year data which has
been made available by the WMAP collaboration. The angular resolution of this
data should allow values of $l$ up to 300-400, but we have chosen to take the
maximum value of $l$ for which detector noise can be neglected, and this is
around $l=300$. We have used 800 values of $r$ in the range $13750{\rm
Mpc}<r<14350{\rm Mpc}$ and 200
values of $r$ for $8750{\rm Mpc}<r<13750{\rm Mpc}$. (Values of $r$ outside
this range appear to make very little change to the estimators).

The values of the estimators and the variance of the ideal experiment are shown
in table \ref{table}. The rms values of the estimators for a set of randomly
generated gaussian maps have been used to check that the linear term in the
estimator removes the effects of the mask without biasing the estimator.
Leaving out the linear term gives a spurious signal.

The results for $f_{CI}$ can be expressed as a
constraint on the more commonly used parameter $f_{NL}$, where
$f_{CI}=0.6f_{NL}$,
\begin{equation}
-26<f_{NL}<109\hbox{~~~~~at 95\% C.L.}
\end{equation}
which we should stress ignores detector noise. This is consistent with the
results published by the WMAP collaboration given the fact that the detector
noise increases the standard deviation of the estimator by 10-20\%
\cite{Komatsu:2003fd}. The results for the WI model are
\begin{equation}
-375<f_{WI}<36.8\hbox{~~~~~at 95\% C.L.}
\end{equation}
Although there is no evidence for a signal, it may be worth pointing out that
$f_{WI}<0$ at 94\% C.L., and this would be in agreement with the predictions of
the warm inflationary model
\cite{Moss:2007cv}.

\begin{table}[ht!]
\begin{center}
\begin{tabular}{|cccccccc|}
\hline
Map&Mask&$f_{\rm sky}$&$l$&$f_{CI}$&$\Delta f_{CI}$&$f_{WI}$&$\Delta f_{WI}$\\
\hline\hline     
W&kp0&0.765&300&24.9&20.5&-169&105\\
V&kp0&0.765&300&33.7&20.5&-174&105\\
R&kp0&0.765&300&-0.712&13.4&3.09&89.2\\
\hline                        
\end{tabular}
\caption{Values of the estimator (with the linear correction) and variance for
a selection of 3-year WMAP datasets and masks. The frequency bands are denoted
by $W$ and $V$ and the maps
are `foreground reduced'. The row labelled $R$ gives the average and rms
average of the estimator over 20 randomly generated  gaussian maps. The
deviation $\Delta f$ in rows $W$ and $V$ is the deviation for an ideal
experiment scaled by the unmasked fraction of the sky  $f_{\rm sky}$. }
\label{table}
\end{center}
\end{table}

\section{Conclusion}

We have compared two different bispectra which represent the non-gaussianity in
the CMB produced by two different types of inflation. The first bispectrum
(CI) is of the local type, which can result from a curvaton field and the
second (WI) was modelled on a result deduced from the theory of warm
inflation. In the second example we found a new, simple form for the estimator
used to quantify the amount of non-gaussianity present in the CMB data. 

The WI bispectrum demands accurate handling of the numerical integrations and
also careful consideration of the effects introduced by masking out part of
the sky. We have introduced a binning technique to improve the accuracy of the
integrations and used an improved estimator to deal with the masked data.

The variance of the estimator for the WI nongaussianity in an ideal experiment
using modes up to $l=500$ is plotted in figure \ref{plotDf}. Beyond $l=500$,
we would have to improve the resolution in the integrations, remove some of
the lower $l$ modes or adopt an alternative computational strategy
\cite{Smith:2006ud}. For example, using modes in the range $9<l<1800$ sets an
upper bound on the deviation of $\Delta f_{WI}<15.9$.

When applied to the WMAP 3-year data we reproduce similar results to previous
work for the CI bispectrum. So far there is no evidence for the WI bispectrum,
and the limits are
\begin{equation}
-375<f_{WI}<36.8\hbox{~~~~~at 95\% C.L.}
\end{equation}
According to the calculation in  ref. \cite{Moss:2007cv}, the strong regime of
warm inflation leads to the WI bispectrum with
\begin{equation}
f_{WI}\approx-18\ln\left(1+{r\over 14}\right)
\end{equation}
where $r$ is a large parameter. It is possible to predict the probability of a
$2\sigma$ signal from the Planck satellite for this model. Using 
$\Delta f_{WI}<15.9$, a $2\sigma$ signal would be obtained with better than
95\% probability if $r>65$.

\appendix

\section{Some properties of masked data}

Consider the situation where we expect a gaussian signal $\sigma(\hat{\bf n})$
but due to unwanted background sources we mask out an area of the sky 
${\cal M}$. The masked spherical harmonic coefficients are
\begin{equation}
a_{lm}=\int d^2\hat{\bf n} 
\,M(\hat{\bf n})\sigma(\hat{\bf n})Y^*_{lm}(\hat{\bf n})
\end{equation}
where $M(\hat{\bf n})=0$ for $\hat{\bf n} \in {\cal M}$ and $M(\hat{\bf n})=1$
otherwise. The covariance matrix of the masked coefficients is defined by
\begin{equation}
C_{l_1m_1l_2m_2}=\langle a_{l_1m_1}a_{l_2m_2}\rangle.
\end{equation}
This can be expressed in terms of the angular power spectrum $C_l$ of the
full-sky coefficients,
\begin{equation}
C_{l_1m_1l_2m_2}=
\sum_{lm}\,C_l\,\int d^2\hat{\bf n}_1 d^2\hat{\bf n}_2
M(\hat{\bf n}_1)M(\hat{\bf n}_2)
Y^*_{l_1m_1}(\hat{\bf n}_1)Y^*_{l_2m_2}(\hat{\bf n}_2)
Y_{lm}(\hat{\bf n}_1)Y_{lm}(\hat{\bf n}_2).\label{cov}
\end{equation}

The main objective of this appendix is to approximate the sum
\begin{equation}
F(\hat{\bf n})=\sum_{l_1l_2m_1m_2}
b_{l_1}b_{l_2}Y_{l_1m_1}(\hat {\bf n})C_{l_1m_1l_2m_2}
Y_{l_1m_1}(\hat {\bf n}),
\end{equation}
where $b_l$ is a slowly varying function of $l$. Using eq. (\ref{cov}),
\begin{equation}
F(\hat{\bf n})=\int d^2\hat{\bf n}_1 d^2\hat{\bf n}_2
M(\hat{\bf n}_1)M(\hat{\bf n}_2)
K(\hat{\bf n},\hat{\bf n}_1)K(\hat{\bf n},\hat{\bf n}_2)
\sum_{lm}C_lY_{lm}(\hat{\bf n}_1)Y_{lm}(\hat{\bf n}_2)\label{Feq}
\end{equation}
where the kernel $K(\hat{\bf n},\hat{\bf n}')$ is given by
\begin{equation}
K(\hat{\bf n},\hat{\bf n}')=
\sum_{lm}b_lY_{lm}(\hat{\bf n})Y_{lm}(\hat{\bf n}').
\end{equation}

The next step is to clarify the sense in which the $b_l$ are slowly varying.
Introduce a new function $b(t)$ by
\begin{equation}
b_l=\int_0^\infty e^{-l(l+1)t}b(t)dt.
\end{equation}
We say that the $b_l$ are slowly varying if $b(t)=0$ for $t>\mu$, where $\mu$
is a small parameter. For primordial fluctuations, the scale of variation of
the $b_l$'s is set by the acoustic oscillations, and $\mu$ is a small
parameter determined by the conformal time at the last scattering surface. (In
reallity $b(t)$ will not vanish for $t>\mu$, but it must be relatively small
in view of the smoothness of the plots of the $b_l$'s).

The kernel $K(\hat{\bf n},\hat{\bf n}')$ is related to the fundamental solution
to the heat equation on a unit sphere $K(\hat{\bf n},\hat{\bf n}',t)$ by
\begin{equation}
K(\hat{\bf n},\hat{\bf n}')=
\int_0^\infty dt \,K(\hat{\bf n},\hat{\bf n}',t)b(t).
\end{equation}
For small time, we can use the asymptotic expansion
\begin{equation}
K(\hat{\bf n},\hat{\bf n}',t)
={1\over 4\pi t}e^{-d(\hat{\bf n},\hat{\bf n}')^2/4t}\left(1+O(t^2)\right),
\end{equation}
where $d(\hat{\bf n},\hat{\bf n}')$ is the great circle distance between 
$\hat{\bf n}$ and $\hat{\bf n}'$, to get an upper bound,
\begin{equation}
|K(\hat{\bf n},\hat{\bf n}')|\le K_0e^{-d(\hat{\bf n},\hat{\bf n}')^2/4\eta}
\label{bk}
\end{equation}
where $K_0=\sum_l(2l+1)b_l$ and $O(\eta^2)$ corrections are dropped.

If the mask is empty, then the orthogonality property of the spherical
harmonics can be used to show that the function $F$ takes a constant value
$S$, where
\begin{equation}
S=\sum_l\,{2l+1\over 4\pi}b_l^2C_l.
\end{equation}
Now use eq. (\ref{Feq}) to calculate the difference
\begin{equation}
| F(\hat{\bf n})-SM(\hat{\bf n})|\le
\int d^2\hat{\bf n}_1 d^2\hat{\bf n}_2
\left|(M(\hat{\bf n}_1)M(\hat{\bf n}_2)-M(\hat{\bf n}))
K(\hat{\bf n},\hat{\bf n}_1)K(\hat{\bf n},\hat{\bf n}_2)
\sum_{lm}C_lY_{lm}(\hat{\bf n}_1)Y_{lm}(\hat{\bf n}_2)\right|
\end{equation}
We have a bound (\ref{bk}) on the kernel and we can bound the final sum by a
constant $C$,
\begin{equation}
\left|\sum_{lm}C_lY_{lm}(\hat{\bf n}_1)Y_{lm}(\hat{\bf n}_2)\right|
\le C=\sum_l {4\pi\over 2l+1}C_l.
\end{equation}
Consider the case where $\hat{\bf n}\in{\cal M}$, which forces the integrand
into the rest of the sky $\bar{\cal M}$,
\begin{equation}
| F(\hat{\bf n})-SM(\hat{\bf n})|\le
CK_0^2\,\left(\int_{\bar{\cal M}} d^2\hat{\bf n}'
e^{-d(\hat{\bf n},\hat{\bf n}')^2/4\eta}\right)^2
\end{equation}
There is an assymptotic expansion for the integral for small $\eta$ obtained by 
setting $\hat{\bf n}'=\hat{\bf n}_c+\epsilon$, where $\hat{\bf n}_c$ is the
closest point on the edge of the mask to $\hat{\bf n}$. The result is that
\begin{equation}
| F(\hat{\bf n})-SM(\hat{\bf n})|\le
CK_0^2\pi(2\eta)^{3}e^{-d(\hat{\bf n},\hat{\bf n}_c)^2/2\eta}.
\end{equation}
A similar argument applies when  $\hat{\bf n}\in\bar {\cal M}$. The
approximation $F\approx SM(\hat{\bf n})$ holds as long as the point on the sky
does not approach too closely the edge of the mask where 
$d(\hat{\bf n},\hat{\bf n}_c)=0$. 

%%%%%%%%%%%%%%%%%%%%%%%%%%%%%%%%%%%%%%%%%%%%%
\bibliography{paper.bib,cosper.bib}

\begin{thebibliography}{25}
\expandafter\ifx\csname natexlab\endcsname\relax\def\natexlab#1{#1}\fi
\expandafter\ifx\csname bibnamefont\endcsname\relax
  \def\bibnamefont#1{#1}\fi
\expandafter\ifx\csname bibfnamefont\endcsname\relax
  \def\bibfnamefont#1{#1}\fi
\expandafter\ifx\csname citenamefont\endcsname\relax
  \def\citenamefont#1{#1}\fi
\expandafter\ifx\csname url\endcsname\relax
  \def\url#1{\texttt{#1}}\fi
\expandafter\ifx\csname urlprefix\endcsname\relax\def\urlprefix{URL }\fi
\providecommand{\bibinfo}[2]{#2}
\providecommand{\eprint}[2][]{\url{#2}}

\bibitem[{\citenamefont{Falk et~al.}(1993)\citenamefont{Falk, Rangarajan, and
  Srednicki}}]{Falk:1992sf}
\bibinfo{author}{\bibfnamefont{T.}~\bibnamefont{Falk}},
  \bibinfo{author}{\bibfnamefont{R.}~\bibnamefont{Rangarajan}},
  \bibnamefont{and}
  \bibinfo{author}{\bibfnamefont{M.}~\bibnamefont{Srednicki}},
  \bibinfo{journal}{Astrophys. J.} \textbf{\bibinfo{volume}{403}},
  \bibinfo{pages}{L1} (\bibinfo{year}{1993}), \eprint{astro-ph/9208001}.

\bibitem[{\citenamefont{Gangui et~al.}(1994)\citenamefont{Gangui, Lucchin,
  Matarrese, and Mollerach}}]{Gangui:1993tt}
\bibinfo{author}{\bibfnamefont{A.}~\bibnamefont{Gangui}},
  \bibinfo{author}{\bibfnamefont{F.}~\bibnamefont{Lucchin}},
  \bibinfo{author}{\bibfnamefont{S.}~\bibnamefont{Matarrese}},
  \bibnamefont{and}
  \bibinfo{author}{\bibfnamefont{S.}~\bibnamefont{Mollerach}},
  \bibinfo{journal}{Astrophys. J.} \textbf{\bibinfo{volume}{430}},
  \bibinfo{pages}{447} (\bibinfo{year}{1994}), \eprint{astro-ph/9312033}.

\bibitem[{\citenamefont{Acquaviva et~al.}(2003)\citenamefont{Acquaviva,
  Bartolo, Matarrese, and Riotto}}]{Acquaviva:2002ud}
\bibinfo{author}{\bibfnamefont{V.}~\bibnamefont{Acquaviva}},
  \bibinfo{author}{\bibfnamefont{N.}~\bibnamefont{Bartolo}},
  \bibinfo{author}{\bibfnamefont{S.}~\bibnamefont{Matarrese}},
  \bibnamefont{and} \bibinfo{author}{\bibfnamefont{A.}~\bibnamefont{Riotto}},
  \bibinfo{journal}{Nucl. Phys.} \textbf{\bibinfo{volume}{B667}},
  \bibinfo{pages}{119} (\bibinfo{year}{2003}), \eprint{astro-ph/0209156}.

\bibitem[{\citenamefont{Komatsu and Spergel}(2001)}]{Komatsu:2001rj}
\bibinfo{author}{\bibfnamefont{E.}~\bibnamefont{Komatsu}} \bibnamefont{and}
  \bibinfo{author}{\bibfnamefont{D.~N.} \bibnamefont{Spergel}},
  \bibinfo{journal}{Phys. Rev.} \textbf{\bibinfo{volume}{D63}},
  \bibinfo{pages}{063002} (\bibinfo{year}{2001}), \eprint{astro-ph/0005036}.

\bibitem[{\citenamefont{Munshi et~al.}(1995)\citenamefont{Munshi, Souradeep,
  and Starobinsky}}]{Munshi:1995eh}
\bibinfo{author}{\bibfnamefont{D.}~\bibnamefont{Munshi}},
  \bibinfo{author}{\bibfnamefont{T.}~\bibnamefont{Souradeep}},
  \bibnamefont{and} \bibinfo{author}{\bibfnamefont{A.~A.}
  \bibnamefont{Starobinsky}}, \bibinfo{journal}{Astrophys. J.}
  \textbf{\bibinfo{volume}{454}}, \bibinfo{pages}{552} (\bibinfo{year}{1995}),
  \eprint{astro-ph/9501100}.

\bibitem[{\citenamefont{Pyne and Carroll}(1996)}]{Pyne:1995bs}
\bibinfo{author}{\bibfnamefont{T.}~\bibnamefont{Pyne}} \bibnamefont{and}
  \bibinfo{author}{\bibfnamefont{S.~M.} \bibnamefont{Carroll}},
  \bibinfo{journal}{Phys. Rev.} \textbf{\bibinfo{volume}{D53}},
  \bibinfo{pages}{2920} (\bibinfo{year}{1996}), \eprint{astro-ph/9510041}.

\bibitem[{\citenamefont{Lyth et~al.}(2005)\citenamefont{Lyth, Malik, and
  Sasaki}}]{Lyth:2004gb}
\bibinfo{author}{\bibfnamefont{D.~H.} \bibnamefont{Lyth}},
  \bibinfo{author}{\bibfnamefont{K.~A.} \bibnamefont{Malik}}, \bibnamefont{and}
  \bibinfo{author}{\bibfnamefont{M.}~\bibnamefont{Sasaki}},
  \bibinfo{journal}{JCAP} \textbf{\bibinfo{volume}{0505}}, \bibinfo{pages}{004}
  (\bibinfo{year}{2005}), \eprint{astro-ph/0411220}.

\bibitem[{\citenamefont{Sasaki et~al.}(2006)\citenamefont{Sasaki, Valiviita,
  and Wands}}]{Sasaki:2006kq}
\bibinfo{author}{\bibfnamefont{M.}~\bibnamefont{Sasaki}},
  \bibinfo{author}{\bibfnamefont{J.}~\bibnamefont{Valiviita}},
  \bibnamefont{and} \bibinfo{author}{\bibfnamefont{D.}~\bibnamefont{Wands}},
  \bibinfo{journal}{Phys. Rev.} \textbf{\bibinfo{volume}{D74}},
  \bibinfo{pages}{103003} (\bibinfo{year}{2006}), \eprint{astro-ph/0607627}.

\bibitem[{\citenamefont{Dvali et~al.}(2004)\citenamefont{Dvali, Gruzinov, and
  Zaldarriaga}}]{Dvali:2003ar}
\bibinfo{author}{\bibfnamefont{G.}~\bibnamefont{Dvali}},
  \bibinfo{author}{\bibfnamefont{A.}~\bibnamefont{Gruzinov}}, \bibnamefont{and}
  \bibinfo{author}{\bibfnamefont{M.}~\bibnamefont{Zaldarriaga}},
  \bibinfo{journal}{Phys. Rev.} \textbf{\bibinfo{volume}{D69}},
  \bibinfo{pages}{083505} (\bibinfo{year}{2004}), \eprint{astro-ph/0305548}.

\bibitem[{\citenamefont{Moss and Xiong}(2007)}]{Moss:2007cv}
\bibinfo{author}{\bibfnamefont{I.~G.} \bibnamefont{Moss}} \bibnamefont{and}
  \bibinfo{author}{\bibfnamefont{C.}~\bibnamefont{Xiong}},
  \bibinfo{journal}{JCAP} \textbf{\bibinfo{volume}{0704}}, \bibinfo{pages}{007}
  (\bibinfo{year}{2007}), \eprint{astro-ph/0701302}.

\bibitem[{\citenamefont{Bardeen et~al.}(1983)\citenamefont{Bardeen, Steinnhart,
  and Turner}}]{bardeen83}
\bibinfo{author}{\bibfnamefont{J.~M.} \bibnamefont{Bardeen}},
  \bibinfo{author}{\bibfnamefont{P.~J.} \bibnamefont{Steinnhart}},
  \bibnamefont{and} \bibinfo{author}{\bibfnamefont{M.~S.}
  \bibnamefont{Turner}}, \bibinfo{journal}{Phys. Rev. D}
  \textbf{\bibinfo{volume}{28}}, \bibinfo{pages}{679} (\bibinfo{year}{1983}).

\bibitem[{\citenamefont{Lyth and Rodriguez}(2005)}]{Lyth:2005fi}
\bibinfo{author}{\bibfnamefont{D.~H.} \bibnamefont{Lyth}} \bibnamefont{and}
  \bibinfo{author}{\bibfnamefont{Y.}~\bibnamefont{Rodriguez}},
  \bibinfo{journal}{Phys. Rev. Lett.} \textbf{\bibinfo{volume}{95}},
  \bibinfo{pages}{121302} (\bibinfo{year}{2005}), \eprint{astro-ph/0504045}.

\bibitem[{\citenamefont{Spergel et~al.}(2006)}]{Spergel:2006hy}
\bibinfo{author}{\bibfnamefont{D.~N.} \bibnamefont{Spergel}}
  \bibnamefont{et~al.} (\bibinfo{year}{2006}), \eprint{astro-ph/0603449}.

\bibitem[{\citenamefont{Moss}(1985)}]{Moss85}
\bibinfo{author}{\bibfnamefont{I.~G.} \bibnamefont{Moss}},
  \bibinfo{journal}{Phys. lett.} \textbf{\bibinfo{volume}{154B}},
  \bibinfo{pages}{120} (\bibinfo{year}{1985}).

\bibitem[{\citenamefont{Berera}(1995)}]{berera95}
\bibinfo{author}{\bibfnamefont{A.}~\bibnamefont{Berera}},
  \bibinfo{journal}{Phys. Rev. lett.} \textbf{\bibinfo{volume}{75}},
  \bibinfo{pages}{3218} (\bibinfo{year}{1995}).

\bibitem[{\citenamefont{Liguori et~al.}(2006)\citenamefont{Liguori, Hansen,
  Komatsu, Matarrese, and Riotto}}]{Liguori:2005rj}
\bibinfo{author}{\bibfnamefont{M.}~\bibnamefont{Liguori}},
  \bibinfo{author}{\bibfnamefont{F.~K.} \bibnamefont{Hansen}},
  \bibinfo{author}{\bibfnamefont{E.}~\bibnamefont{Komatsu}},
  \bibinfo{author}{\bibfnamefont{S.}~\bibnamefont{Matarrese}},
  \bibnamefont{and} \bibinfo{author}{\bibfnamefont{A.}~\bibnamefont{Riotto}},
  \bibinfo{journal}{Phys. Rev.} \textbf{\bibinfo{volume}{D73}},
  \bibinfo{pages}{043505} (\bibinfo{year}{2006}), \eprint{astro-ph/0509098}.

\bibitem[{\citenamefont{Smith and Zaldarriaga}(2006)}]{Smith:2006ud}
\bibinfo{author}{\bibfnamefont{K.~M.} \bibnamefont{Smith}} \bibnamefont{and}
  \bibinfo{author}{\bibfnamefont{M.}~\bibnamefont{Zaldarriaga}}
  (\bibinfo{year}{2006}), \eprint{astro-ph/0612571}.

\bibitem[{\citenamefont{Creminelli et~al.}(2006)\citenamefont{Creminelli,
  Nicolis, Senatore, Tegmark, and Zaldarriaga}}]{Creminelli:2005hu}
\bibinfo{author}{\bibfnamefont{P.}~\bibnamefont{Creminelli}},
  \bibinfo{author}{\bibfnamefont{A.}~\bibnamefont{Nicolis}},
  \bibinfo{author}{\bibfnamefont{L.}~\bibnamefont{Senatore}},
  \bibinfo{author}{\bibfnamefont{M.}~\bibnamefont{Tegmark}}, \bibnamefont{and}
  \bibinfo{author}{\bibfnamefont{M.}~\bibnamefont{Zaldarriaga}},
  \bibinfo{journal}{JCAP} \textbf{\bibinfo{volume}{0605}}, \bibinfo{pages}{004}
  (\bibinfo{year}{2006}), \eprint{astro-ph/0509029}.

\bibitem[{\citenamefont{Creminelli}(2003)}]{Creminelli:2003iq}
\bibinfo{author}{\bibfnamefont{P.}~\bibnamefont{Creminelli}},
  \bibinfo{journal}{JCAP} \textbf{\bibinfo{volume}{0310}}, \bibinfo{pages}{003}
  (\bibinfo{year}{2003}), \eprint{astro-ph/0306122}.

\bibitem[{\citenamefont{Seljak and Zaldarriaga}(1996)}]{Seljak:1996is}
\bibinfo{author}{\bibfnamefont{U.}~\bibnamefont{Seljak}} \bibnamefont{and}
  \bibinfo{author}{\bibfnamefont{M.}~\bibnamefont{Zaldarriaga}},
  \bibinfo{journal}{Astrophys. J.} \textbf{\bibinfo{volume}{469}},
  \bibinfo{pages}{437} (\bibinfo{year}{1996}), \eprint{astro-ph/9603033}.

\bibitem[{\citenamefont{Brink and Satchler}(1993)}]{brink}
\bibinfo{author}{\bibfnamefont{D.~M.} \bibnamefont{Brink}} \bibnamefont{and}
  \bibinfo{author}{\bibfnamefont{G.~R.} \bibnamefont{Satchler}},
  \emph{\bibinfo{title}{Angular Momentum}} (\bibinfo{publisher}{Oxford Science
  Publications}, \bibinfo{year}{1993}), \bibinfo{edition}{3rd} ed.,
  \bibinfo{note}{appendix II}.

\bibitem[{\citenamefont{Komatsu}(2002)}]{Komatsu:2002db}
\bibinfo{author}{\bibfnamefont{E.}~\bibnamefont{Komatsu}}
  (\bibinfo{year}{2002}), \eprint{astro-ph/0206039}.

\bibitem[{\citenamefont{Babich et~al.}(2004)\citenamefont{Babich, Creminelli,
  and Zaldarriaga}}]{Babich:2004gb}
\bibinfo{author}{\bibfnamefont{D.}~\bibnamefont{Babich}},
  \bibinfo{author}{\bibfnamefont{P.}~\bibnamefont{Creminelli}},
  \bibnamefont{and}
  \bibinfo{author}{\bibfnamefont{M.}~\bibnamefont{Zaldarriaga}},
  \bibinfo{journal}{JCAP} \textbf{\bibinfo{volume}{0408}}, \bibinfo{pages}{009}
  (\bibinfo{year}{2004}), \eprint{astro-ph/0405356}.

\bibitem[{\citenamefont{Komatsu et~al.}(2005)\citenamefont{Komatsu, Spergel,
  and Wandelt}}]{Komatsu:2003iq}
\bibinfo{author}{\bibfnamefont{E.}~\bibnamefont{Komatsu}},
  \bibinfo{author}{\bibfnamefont{D.~N.} \bibnamefont{Spergel}},
  \bibnamefont{and} \bibinfo{author}{\bibfnamefont{B.~D.}
  \bibnamefont{Wandelt}}, \bibinfo{journal}{Astrophys. J.}
  \textbf{\bibinfo{volume}{634}}, \bibinfo{pages}{14} (\bibinfo{year}{2005}),
  \eprint{astro-ph/0305189}.

\bibitem[{\citenamefont{Komatsu et~al.}(2003)}]{Komatsu:2003fd}
\bibinfo{author}{\bibfnamefont{E.}~\bibnamefont{Komatsu}} \bibnamefont{et~al.}
  (\bibinfo{collaboration}{WMAP}), \bibinfo{journal}{Astrophys. J. Suppl.}
  \textbf{\bibinfo{volume}{148}}, \bibinfo{pages}{119} (\bibinfo{year}{2003}),
  \eprint{astro-ph/0302223}.

\end{thebibliography}

\end{document}